# Current potentials and challenges using Sentinel-1 for broadacre field remote sensing

**Martin Peter Christiansen [a,c,\*], Morten Stigaard Laursen[a], Birgitte Feld Mikkelsen[b], Nima Teimouri[a], Rasmus Nyholm Jørgensen[a], Claus Aage Grøn Sørensen[a]**

[a] Department of Engineering, Aarhus University, Finlandsgade 22, 8200 Aarhus N, Denmark

[b] AgroTech, Danish Technological Institute, Agro Food Park 15, 8200 Aarhus N, Denmark

[c] SDU UAS Center, University of Southern Denmark, Campusvej 55, 5230 Odense M, Denmark

* Corresponding author. Email: mapc@mmmi.sdu.dk

## Abstract

ESA operates the Sentinel-1 satellites, which provides Synthetic Aperture Radar (SAR) data of Earth. Recorded Sentinel-1 data have shown a potential for remotely observing and monitoring local conditions on broad acre fields. Remote sensing using Sentinel-1 have the potential to provide daily updates on the current conditions in the individual fields and at the same time give an overview of the agricultural areas in the region. Research depends on the ability of independent validation of the presented results. In the case of the Sentinel-1 satellites, every researcher has access to the same base dataset, and therefore independent validation is possible. Well documented research performed with Sentinel-1 allow other research the ability to redo the experiments and either validate or falsify presented findings. Based on current state-of-art research we have chosen to provide a service for researchers in the agricultural domain. The service allows researchers the ability to monitor local conditions by using the Sentinel-1 information combined with a priori knowledge from broad acre fields. Correlating processed Sentinel-1 to the actual conditions is still a task the individual researchers must perform to benefit from the service. In this paper, we presented our methodology in translating sentinel-1 data to a level that is more accessible to researchers in the agricultural field. The goal here was to make the data more easily available, so the primary focus can be on correlating and comparing to measurements collected in the broadacre fields. We illustrate the value of the service with three examples of the possible application areas. The presented application examples are all based on Denmark, where we have processed all sentinel-1 scan from since 2016.

## 1. Introduction

Free Big data have become available in recent years from different remote sensing sources. Both optical and synthetic aperture radar (SAR) remote sensing solutions are possible such as the Sentinel-1 (Torres et al. 2012), and Sentinel-2 (Drusch et al. 2012) satellites (ESA, Paris, France) within the European Copernicus program and the Landsat-7 (Roy et al. 2016) and Landsat-8 (Roy et al. 2014) provided in cooperation between the National Aeronautics and Space Administration and U.S. Geological Survey. Data from these different remote sensing satellite solutions are freely available and have the potential to open up a broad spectrum of monitoring and prediction applications for the agricultural domain (Dalla Mura et al. 2015), (Kussul et al. 2017).

The Sentinel-1 satellite data have already been used in many articles related to the agricultural domain with interesting results (Erten et al. 2016), (Zhou, Li, and Pan 2018), (Jin et al. 2018), (Kumar et al. 2017). The Sentinel-1 data over agricultural fields have been used to estimate parameters such as crop type, plant height, biomass, Green Area Index (GAI) and vegetation water content.

Scientific research depends on the ability to perform an independent validation of the presented results (Sagan 2011). In the case of the Sentinel-1 satellites, every researcher have access to the same base data set and therefore independent validation is possible. Well documented research performed with Sentinel-1 remote sensing data will allow other research the ability to redo the experiments and either validate or falsify findings.

To estimate different crop parameters, one requires reference data from actual fields or a priori created knowledge from literature and research projects. The demands from agricultural to Sentinel-1 and remote sensing satellites in general means reference data is highly relevant for the development process of estimation methods. Public available reference data provides a way to ensure researchers in the same field have the chance to redo the processing and validate the results.

Another challenge is the level of understanding one needs to have in pre-processing sentinel-1 data (Pierdicca, Pulvirenti, and Pace 2014), before an estimation and correlation in an agricultural context can be performed. The needed level of understanding for Sentinel-1 pre-processing creates a need to provide well-documented steps for agricultural researchers to follow. Well documented pre-processing step ensures data are readily available for experts outside the SAR research field and also allows reproduction on the same experiments to either falsify or validate previously created results. The challenge is the number of details one needs to document of these satellite-based systems, to ensure other researchers can reproduce the results.

Based on current research and evaluated results, we currently provide a service for researchers in the agricultural domain. The service allows researchers in the agricultural domain the ability to monitor local conditions by using the recorded sentinel-1 information combined with a priori knowledge from broad acre fields. Correlating the possessed sentinel-1 to actual conditions is still a task the individual researchers must perform to utilize the service.

## 2. Materials

2.1. Satellite data sources

Different free data sources are available for remote sensing with an agricultural use case (Roy et al. 2014), (Li, Jiang, and Feng 2014), (Bosch et al. 2016), (Berger et al. 2012). Currently, only the ESA's Sentinel satellite missions continually provide free sensory data for both multispectral and SAR images via the Copernicus data hub. From the Copernicus open access data hub (Müller et al. 2016), (Klein et al. 2017), one can request and download Sentinel- 1, 2 and 3 products recorded over a specific region. The Sentinel data products need to be processed using software such as SNAP before one can utilize it for a particular purpose.

*2.1.1. Sentinel-1*

Two products created from Sentinel-1 recordings has been analyzed in the agricultural domain, which is Single Look Complex (SLC), and Ground Range Detected (GRD) SAR data (Torres et al. 2012). Significant parts of the current research for the agricultural domain using Sentinel-1 data have been focusing on GRD since it comes more preprocessed and has four to five times smaller package size compared SLC. The GRD standard products come as either VV+VH or HH+HV, with a swath of 250km (Torres et al. 2012). On a global scale where broadacre crops are grown, Sentinel-1 mainly records data for the VV or VV+VH polarizations. For the growth season 2016/17 in Denmark, Sentinel GRD VV+VH data (dual polarisation) are available, from the two sentinel-1 satellites, with one to three days in between. The GRD products come in both ascending and descending recording mode and from different satellite tracks.

2.2. External services to combine with the Sentinel data

In agriculture, Sentinel-1 satellite data generally needs to be combined with a prior knowledge about SAR measurements to land and vegetation, or with models created by correlating local observations from broad acre fields with the SAR measurements. In this section, we present the different external data sources relevant when analysing Sentinel-1 data.

*2.2.1. Land-parcel identification system*

The Land-parcel identification system (LPIS) can be used retrieve information about active fields for each season and what the individual applicants are currently growing. The applicant is the farmer which has requested funding from the European Agricultural Guarantee Fund (EAGF) for a specific field. The field boundaries are drawn manually, by the individual applicant, on orthophotos recorded with planes over Denmark. The farmers are expected to exclude unfarmed land and ineligible features from the recorded field boundaries (Phil Wynn Owen et al. 2016). According to the EAGF, the estimated error level in LPIS was 2.9 % in 2014, and close to half was related to the marking of the correct area. An error level 2.9%, makes LPIS data an interesting reference source and annotated dataset for analysing Sentinel data from individual fields. Combining Sentinel-1 time series and LPIS data could be a future means to track and predict errors in the data supplied by the individual farmer (Giasat s.r.o 2017).

*2.2.2. Nordic field trial system*

Nordic Field Trial System (NFTS) is a repository provided by SEGES and Danish Technological Institute to design experiments plans and store recorded data from field trials (Hansen et al. 2004). Experiments from Denmark, Norway, and Sweden have been recorded using NFTS since 2006. The NFTS repository is a source of reference datasets with fields spread out all over Denmark and the other Nordic countries. NFTS is an open access repository that provides the public with the ability to retrieve broadacre field experiment plans and results via their homepage. Growth stage, crop treatment, sensor recordings, yield estimates for different crop types are some of the relevant parameters that can be extracted from the NFTS repository. Globally alternative data sources exist such as the AgTrials repository that attempts to use a standardized definition from the Crop Ontology for all measurements (Hyman et al. 2017).

*2.2.3. Fieldbabel service*

The Fieldbabel service is intended to allow agricultural researchers the ability to rapidly compare and correlate data collected in the field with remote sensing measurement from satellites. The fieldbabel service is built on open-source software to handle preprocessing of the request from the users and export the results as QGIS projects (Shelestov et al. 2013). Commercial alternatives exist to QGIS, but in this case, we are attempting to keep the source code and functionality open to researchers without the need to pay additional fees for using the service.

Sentinel Application Platform (SNAP) is the standard software tool provided by ESA (Zuhlke et al. 2015), jointly developed by Brockmann Consult, Array Systems Computing and C-S, to process data from the Sentinel satellites. The Fieldbabel service is based upon the SNAP API and tools from ESA to process and extract the relevant area.

Python 2.7 running on an Ubuntu 16.04 server is used for both the frontend and backend parts of the presented service. The frontend is based on the python Flask API (Grinberg 2018) which is a micro web framework and template development. Shapefiles added to the Qgis project are handled using python Shapefile Library (Butler 2005), which can be used to process, store and exchange GIS information.

### 3. Methods

### 3.1. Sentinel-1 GRD preprocessing pipeline

In this section, we present the processing steps that are used on the field babel service for preprocessing the different sentinel satellite data sources. The preprocessing pipeline performs the procedure illustrated in the SNAP graph in Figure 1.

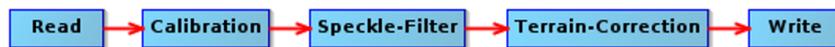

Figure 1. The preprocessing stages used on the fieldbabel server. The raw GRD are read, processed in the SNAP toll and the result saved on the filedbabel server

The GRD product is loaded with the SNAP tool and calibrated to the sigma nought area on the ground (Sabel et al. 2012), providing a layer in which the values are directly related to the radar backscatter (Sabel et al. 2012). To remove noise, a Lee sigma speckle filter (Lee et al. 2009) with a single look; a window size of 7x7; a sigma value of 0.9 and a target window size of 3x3 is used. The terrain correction operation is intended to compensate for distortions so the geometry of the layer will be as close as possible to the actual world. For terrain correction, the pipeline utilizes the Range doppler terrain correction (Schubert et al. 2015) with the digital elevation model parameter set to "SRTM 1Sec. HGT" and the pixel spacing of 10m.

### 3.2. Fieldbabel processing of requests

The Fieldbabel server follows the processing flow illustrated in Figure 2. The processes are split up into two stages; one is the web-service that handles the user requests directly, and the second is the actual processing and creation of the Qgis project.

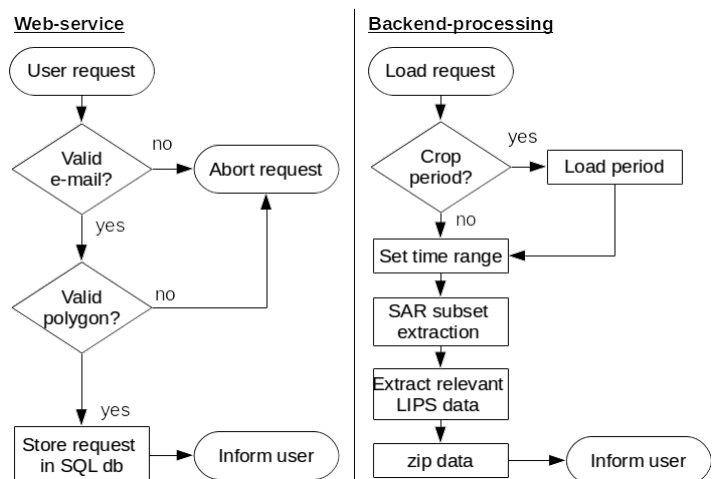

Figure 2.  Flowchart of the fieldbabel server processing of requests.

#### 3.2.1. Web-service

The web-service is designed to handle user-requests for their selected areas of Denmark. A Geojson polygon and an e-mail are submitted and evaluated via the web interface. If the e-mail address is valid and the geojson contains a single polygon with an area no larger than a single degree latitude/longitude, the request is stored in the SQL database on the server and set for processing.

#### 3.2.2. Backend-processing

The processing is performed using a background process running on the Linux server. The server loads the relevant period for the selected area. The server fetches the relevant preprocessed sentinel data for the selected period and area and exports the layers as GeoTIFF's. The backend processing utilises the layers created with the preprocessed pipeline, by extracting a subset of the layer and converting the VV and VH values into dB scale. Based on the Sentinel-1 output from the preprocessing pipelines, the GeoTIFF values for the red, green and blue bands are represented by VV(dB) and VH(dB) and VV(dB)/VH(dB) respectively.

#### 3.2.3. Qgis project creation

The GeoTIFF files are packaged as a Qgis project for the request with preset color ranges for each band. The visual color range for the GeoTIFF in Qgis is set by having analyzed the value spectrum of all layers over a whole year,  in a specific track of Sentinel-1 scans from Denmark. The relevant LPIS data for the same period and area is added to the Qgis project as a final step before being compressed into a zip file. When the Qgis project have been packed into a zip file, the user is informed with a link for the project to download.

### 3.3. Crop selection for analysis

Information about the start and end of a crops regular growing season can be used to track the temporal behavior in the recorded Sentinel-1 data. Local weather and geographic specific conditions impact the growing season for any given area, making it impossible to provide exact dates. Different crops are seeded and harvested at particular periods of the year, for Nordic countries such as Denmark, since they are seasonally dependent. Information about crop seasons is highly relevant for researchers outside the agricultural domain, such as engineers with images and signal processing backgrounds. The temporal information about the crop ensures researchers the ability to analyse the behavior when they are active. By researching the normal seeded and harvesting for different crop types, this information has been added to the web-service. The field babel service can extract the relevant Sentinel-1 scans for the chosen period, based on the selected crop type.

The primary source for determining crop seeding and harvesting time is the homepage https://dyrk-plant.dlbr.dk/, which is a guide for Danish farmers. Since local wheat and other external factors still impact the growing season, the time-period has been extended with a month in each end. The current stored crop period is documented in Table 1, where xx in the date representing the year. A plus or minus year value indicates if the crops season is part of multiple years.

Table 1. Table with chosen crop type, seasonal start and end date, when extracting data from the pre-processed Sentinel-1 layers on the Fieldbabel server

| Danish name (LPIS entry) | English name (Fieldbabel) | start date | end date |
|---|---|---|---|
|  | All | 01-01-20xx | 31-12-20xx |
| Majs | Corn | 15-03-20xx | 15-11-20xx |
| Vårbyg | Spring barley | 01-03-20xx | 01-09-20xx |
| Sukkerroer | Sugar beat | 01-04-20xx | 01-02-20xx(+1) |
| Vårraps | Spring rape | 01-03-20xx | 01-10-20xx |
| Vinterraps | Winter rapeseed | 01-07-20xx(-1) | 01-08-20xx |
| Vinterhvede | Winter wheat | 15-08-20xx(-1) | 01-10-20xx |

### 3.4. Case studies

#### 3.4.1. Crop-type distinction

Classification and mapping of crop-types have been a common focus for remote sensing (Lopez-Sanchez and Ballester-Berman 2009) for the last decade. The sensitivity of C band SAR data to crop type different features have been validated in (Picard, Le Toan, and Mattia 2003). A number of articles have been published, which indicate that sentinel-1 data can be used to distinguish crop types, by training based on known reference data (Skriver et al. 2011).

Data from the LPIS database can be a means to validate the Qgis projects created with Fieldbabel, by determining if different crop types are visually distinguishable. For this paper an area with an bounding box of (WGS84: latitude$_{min}$ 55.2025, longitude$_{min}$ 10.3275 and latitude$_{max}$ 55.0700, longitude$_{max}$ 10.5839), was chosen from the 1st  of june 2017 where the significant crop types in Denmark are in their late stages of development with the exception of maize and beets (Olesen et al. 2012).

#### 3.4.2. Sampling of field data

Collecting reference data from broadacre fields to correlate Sentinel-1 measurements against can be a time-consuming and labor-intensive task. The goal in collecting reference data is not to cover each field meter by meter, but to acquire samples from larger areas where VV or VH is reporting variation compared to the rest of the field. One way to perform this automatically would be to download the newest Sentinel-1 datasets and cluster areas with high and low values for either VV or VH.

In this example, a field analysed by shrinking the LPIS field boundary with 30m using erosion. The intention with the erosion step is only to allow crop values to be part of the time-temporal series, by shrinking the inner area. The 30m value is chosen based on three factors; an assumption that each processed layer have maximum alignment error of ±10m; the headland of the field is the least consistently seeded; since the farmers manually draw the LPIS field, erosion of the boundaries ensures the processing avoids input from minor manual drawing inaccuracies.

To determine where a sample should be collected, a K-mean clustering algorithm for grids is used to evaluate the distribution of measurements. The K-mean algorithm is set to search with three cluster bins, based on an approach that combines hill-climbing (Wu et al. 2008) and iterative minimum distance (Forgey 1965). The number of cluster bins is set to three to evaluate the homogeneity of the field and cluster the measurements into the main areas of variation. For this paper,  fields with only a few meters of variation in elevation is select and analysed, using the k-mean clustering method.

#### 3.4.3. Phenological stages

The web-service allows researchers the ability to monitor the full crop growth by obtaining timely information about the crop condition regarding sentinel-1 data. According to literature (Veloso et al. 2017) the VV/VH profile are in good agreement with the GAI for winter wheat. GAI is measured as the ratio of green material in the ground area it covers.

The GAI index is known to follow the growth stages of the crop. According to (Gerry Boyle et al. 2016) the GAI will peak between growth stage 39-59 o for winter wheat. This make is interesting to attempt to compare growth stage and the VV(dB)/VH(dB) ration as a time series, to validate if the same peak scenario is present. If VV(dB)/VH(dB) ratio follows the GAI it could indicate sentinel-1 data can be used to provide estimates of the current growth stage of winter wheat.

For trial 010811616 in the NFTS database, the progress of winter wheat growth stage has been noted throughout the growing season. In trial 010811616, there exist five fields scattered throughout Denmark, where the same experiment was performed. Other trails in the NFTS have also logged the growth stages of winter wheat and other crops, but trial 010111616 was the only one found where notation was done consistently for all fields.

## 4. Results and Discussion

The result resection focuses on a number of example cases to illustrate what the service can be used for in the agricultural domain. The results are based on a combination of literature study and experiments using the collected data.

### 4.1. Crop type distinction based on Qgis data

The different crop-field types in the Sentinel-1 image processed with the per-processing pipeline can be seen in Figure 3, which was recorded on the first of june 2017. Fields with maize, spring barley and winter rape are directly distinguishable in Figure 3, since the color patterns differ significantly. Winter barley and wheat seems to have identical structures in the produced Qgis image. The similarity between winter barley and wheat is expected since they would look identical in the 10 m pixel level this time of year. One should note the arrow in Figure 3, where a field seems to contain both winter rape and winter wheat. The marked area might indicate an error made by the farmer when drawing the field boundary.

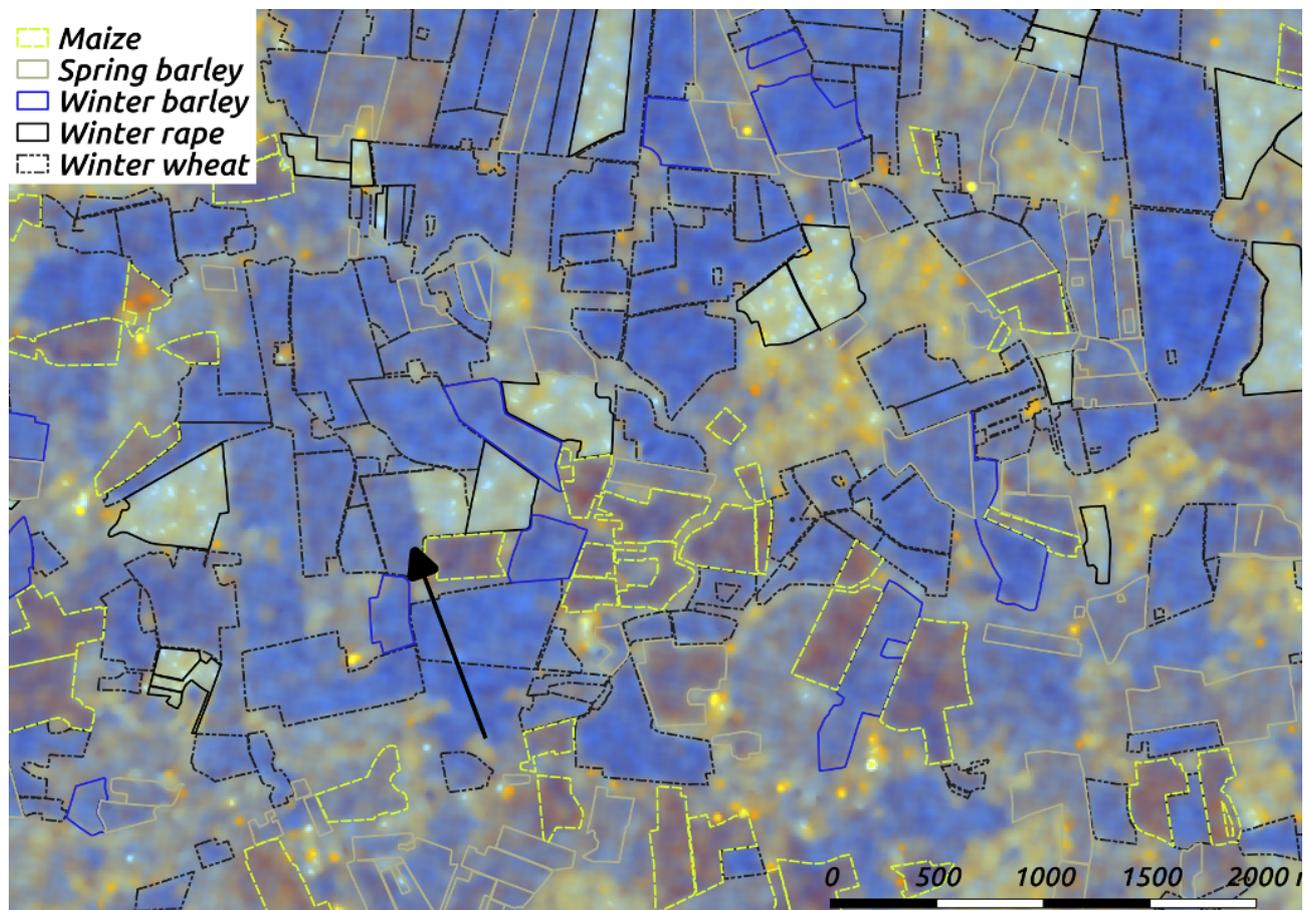

Figure 3. Processed SAR image from the 1st of june 2017 with marking of the five crop types: Maize, Spring barley, Winter barley, Winter rape, Winter wheat.

The achieved result indicates the processed sentinel-1 images provided by the fieldbabel service can be used to identify different crop types, with the exception of winter barley and wheat. By automatically tracking the different crop types in the Sentinel-1 images, since the start of the year, we would expect one could also distinguish winter barley and wheat.

4.2. Sampling in the field based on current Sentinel-1 data
The results of the k-means clustering can be seen in Figure 4. Since the crop field is somewhat flat, the clustering should not be significantly affected by the elevation.

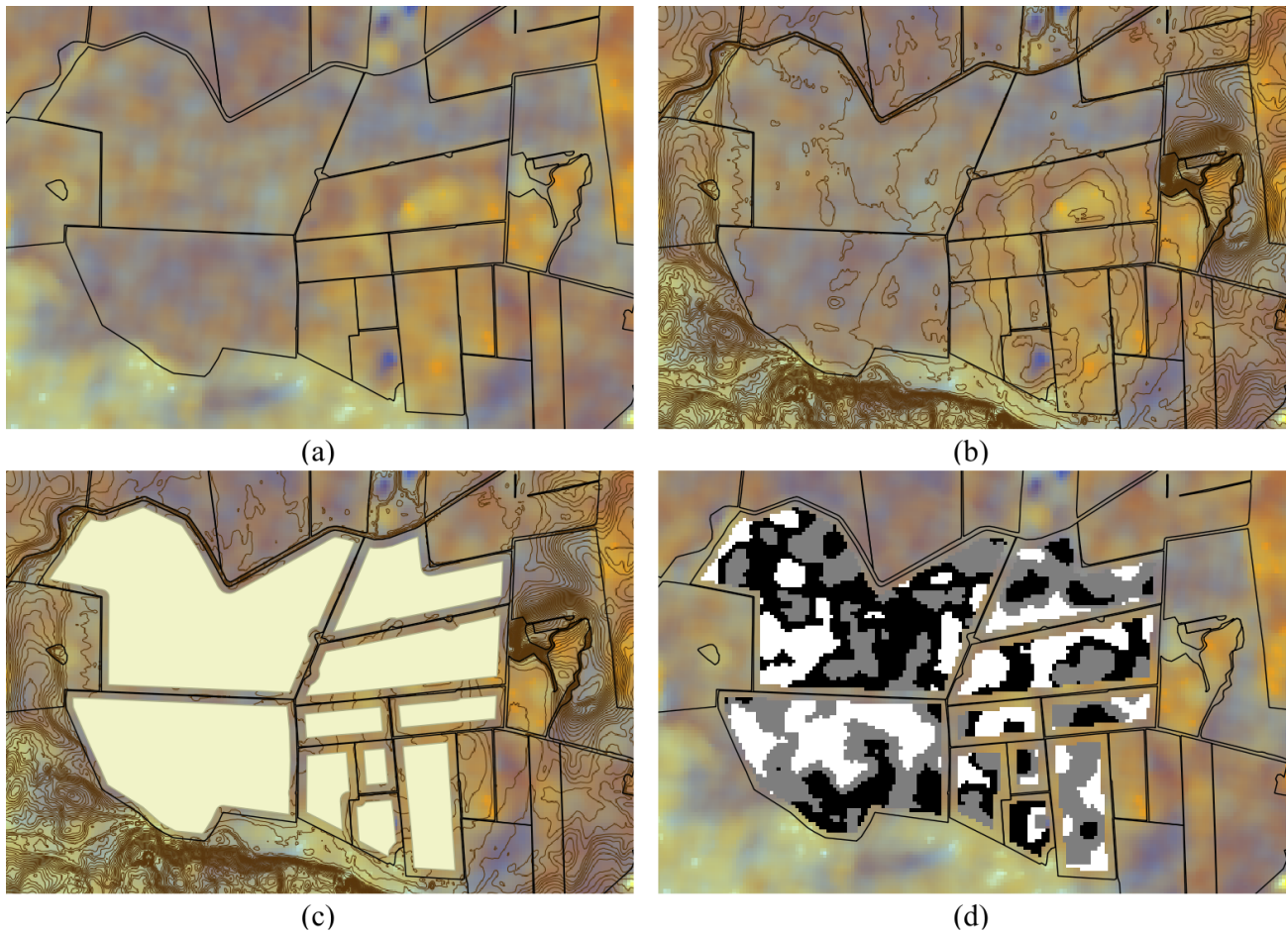

Figure 4.  The processing of individual fields using k-means clustering, to create maps of where to collect samples. The red lines in subfigure (b) represents elevation differences of more than 0.5 meters. (c) Erosion is used to shrink the field boundary.(d) Each field is analysed separately using k-means clustering.

From the above-mentioned figure on can see that the clusters do not seem to follow the evaluation on the area. This approach could be used to plan sampling reference data from a field for both manual collection and more automated systems such as a drone.

4.3. Time series analysis: Phenological stages

After processing trough farm report manager, the time-temporal plot results for the VV(dB)/VH(dB) ratio can be seen in Figure 5 for two of the five fields.

Other trials in the Nordic field trial data have also logged the growth stages of winter wheat and other crops, but trial 010111616 was the only one found where notation was done consistently for all fields. From the growth stages compared to the ratio of VV(dB)/VH(dB) it can be seen, that the curve starts to rise around stage 24-25 and peaks around growth stage 39. The peaking spot on VV(dB)/VH(dB) curve show a tendency similar to GAI for winter wheat, which we would be expected based on earlier studies (Veloso et al. 2017). The dataset indicates a tendency, we would expect generally to be present for winter wheat fields where the correct Sentinel-1 data products are recorded.  The example with winter wheat growth stages and Sentinel-1 data also match the previous findings published by other researchers.

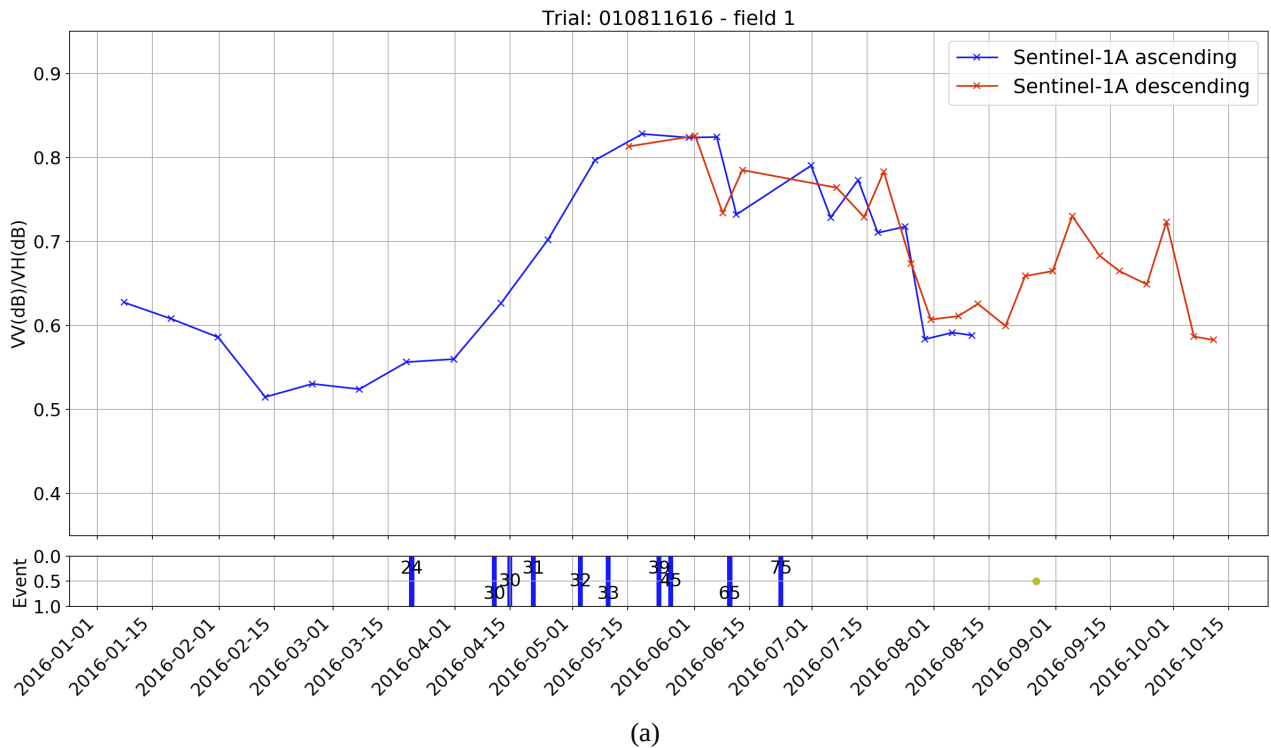

(a)

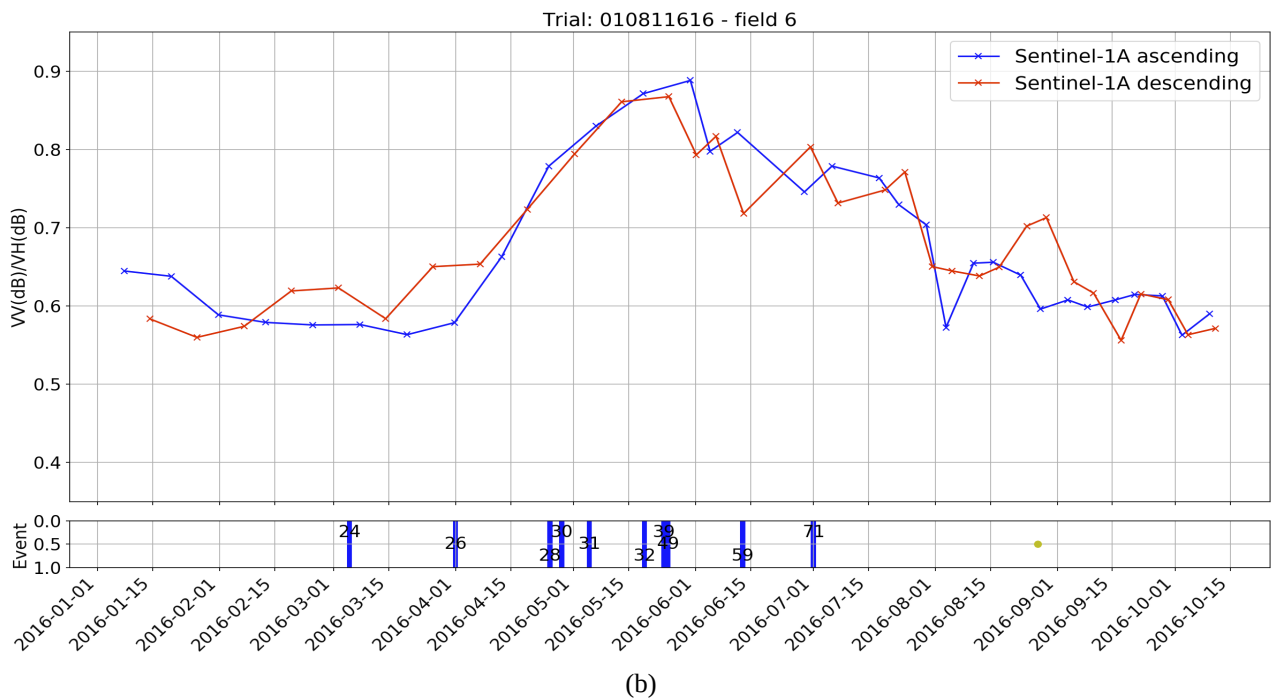

(b)

Figure 5.  Plot of the mean VV/VH value for two specific winter wheat field against time. Below is the reported growth stage of these winter wheat in NFTS.

From a big data perspective automatically determining growth stages of winter winter could be a means to automatically plan field operations such as fertilizing, spraying and harvesting. Automated prediction of winter wheat growth stages could be evaluated on upcoming seasons and will also be explored in the future.

### 5. Conclusions

In this paper, we have presented our Fieldbabel system for pre-processing Sentinel-1 recording for the agricultural domain. The system is intended for non-SAR processing experts, so focus can be on correlation with field samples. Three different application examples have been presented to illustrate the possible potential of the Fieldbabel service.

By providing both application examples and the service, we hope it the will provide more agricultural researchers with the ability to discover new uses for Sentinel-1 data.

## Acknowledgements

The work presented here is supported by the FutureCropping and SqMFarm projects funded by Innovation Fund Denmark and GUPD, The Danish AgriFish Agency, respectively. Data From the National Field Trials, from SEGES L&F are used for ground truth.